# Embedding-based neural network for investment return prediction

Jianlong Zhu[1]
The University of Ballarat,
Ballarat, Australia,
medbsb@163.com,

Fengxiao[3],
Zhejiang Gongshang University,
Hangzhou, Zhejiang Province, China,
ffx562590763@163.com

Dan Xian[2],
Northeastern University
San jose, United States,
xian.d@northeastern.edu

Yichen Nie*,
University of Michigan, Ann Arbor,
Michigan, United States,
yichenn@umich.edu,

*Abstract*—In addition to being familiar with policies, high investment returns also require extensive knowledge of relevant industry knowledge and news. In addition, it is necessary to leverage relevant theories for investment to make decisions, thereby amplifying investment returns. A effective investment return estimate can feedback the future rate of return of investment behavior. In recent years, deep learning are developing rapidly, and investment return prediction based on deep learning has become an emerging research topic. This paper proposes an embedding-based dual branch approach to predict an investment's return. This approach leverages embedding to encode the investment id into a low-dimensional dense vector, thereby mapping high-dimensional data to a low-dimensional manifold, so that high-dimensional features can be represented competitively. In addition, the dual branch model realizes the decoupling of features by separately encoding different information in the two branches. In addition, the swish activation function further improves the model performance. Our approach are validated on the Ubiquant Market Prediction dataset. The results demonstrate the superiority of our approach compared to Xgboost, Lightgbm and Catboost.

## I. INTRODUCTION

The financial market is a complex system, but there are some universal laws hidden in the data. By exploring the statistical characteristics of some observed variables of investment behavior and mining the statistical laws behind them, we can mine the relationship between investment and returns, and then find the operating laws behind them. Deep neural network is essentially a highly complex nonlinear system, so it can identify complicated features in sample information, and has unprecedented high performance in feature extraction. Analyzing from the underlying theory The return on investment is likely to be non-linear, because the return on investment is influenced by multiple pieces of information. Models built using deep neural network technology can capture complex feature associations to more accurately predict return on investment.

This paper will leverage embedding and deep neural network to build a statistical model, conduct a detailed analysis of the complex nonlinear and asymmetric investment and return relationship, and reveal the potential mechanism and law of investment and return, which helps investment institutions make more favorable investment decisions. The Embedding-based dual branch network proposed in this paper is mainly divided into two branches. One branch encodes the investment id into a low-dimensional dense vector by encoding the embedding, thereby mapping the high-dimensional data to the low-dimensional manifold. The second branch re-encodes dense features through a multilayer perceptron. The two branches are combined by feature encoding to further improve the expressiveness of features. In addition, the activation function of each layer of the multilayer perceptron utilizes swish to further improve the performance. The approach proposed in this paper is validated on the Ubiquant Market Prediction dataset, and compared with popular methods such as Lightgbm and Catboost, the experiments proved that our approach achieves more competitive performance.

In summary, the main contributions of this paper are divided into two parts:

● A denser multi-dimensional representation is carried out through embedding, and the calculation process of reward prediction is optimized through the expression of low-dimensional manifold.

● A novel dual branch network is proposed, which combines multi-layer perceptron and feature encoding optimization method to further improve the diversity of feature expression.

## II. RELATED WORK

Financial market-related forecasting has always been a significant field of inquiry in the industry and academia. Especially in recent years, the stock market has been turbulent, and many scholars have devoted themselves to the predictive research of financial stock markets.

Chen et al. proposed a basic hybrid framework of feature-weighted support vector machines and feature-weighted K-nearest neighbors for predicting stock market indices. The framework estimates the significance of each feature by calculating the information gain, and then obtains the weighting coefficient [1]. Chung et al. proposed a method of integrating long short-term memory network and genetic algorithm, leveraging GA to determine the time window size and topology of LSTM network, and then study the time characteristics of stock market [2]. Hu et al. proposed an improved sine cosine algorithm (ISCA) that optimizes the network weights by introducing additional parameters, and leveraged the approach to predict the S&P 500 and the Dow Jones Industrial Average [3]. Lee et al. proposed an end-to-end stock market prediction framework NuNet, which consists of two feature extractors, including a high-dimensional feature extractor and a target-indexed feature extractor. And then propose mini-batch sampling and column-wise random shuffling methods for data augmentation. The approach is validated on three indices [4]. Chen et al. proposed a stock price trend prediction model based on an encoder-decoder framework, which can adaptively predict stock price movements and their durations. This method utilizes an encoder-decoder framework based on a dual attention mechanism to predict stock price trends [5]. Cho et al. proposed a model for investment decision-making in the lending market, which consists of an instance-based entropy-fuzzy support vector machine. Experiments show that the method outperforms six state-of-the-art models [6]. Lu et al. proposed CNN-BiLSTM-AM to predict stock closing prices. The role of CNN in this framework is to extract features. BiLSTM is utilized for time series analysis, and AM is utilized to capture fine-grained features to improve prediction accuracy [7]. Lyu et al. conducted venture capital prediction through graph neural networks, designed an incremental representation learning mechanism and a sequential learning model to optimize the network structure, and achieved competitive performance on a comprehensive dataset of global venture capital [8]. Tsai et al. used a fusion strategy to build an efficient model for predicting stock returns. This method comprehensively considers the fusion method of stacking and bagging. The fusion method which takes the output of several models can significantly improve the performance compared to a single classifier [9,10,11,12,13].

The rest of this paper is as follows. The second section mainly introduces the specific principles and embedding details of the dual branch network. The third section mainly introduces the results and analysis of experiments with dual branch network and embedding optimization methods. The fourth section is the summary of this paper.

## III. METHOD

In this paper, we propose a novel predict investment return Network, the network structure is shown in Figure 1.

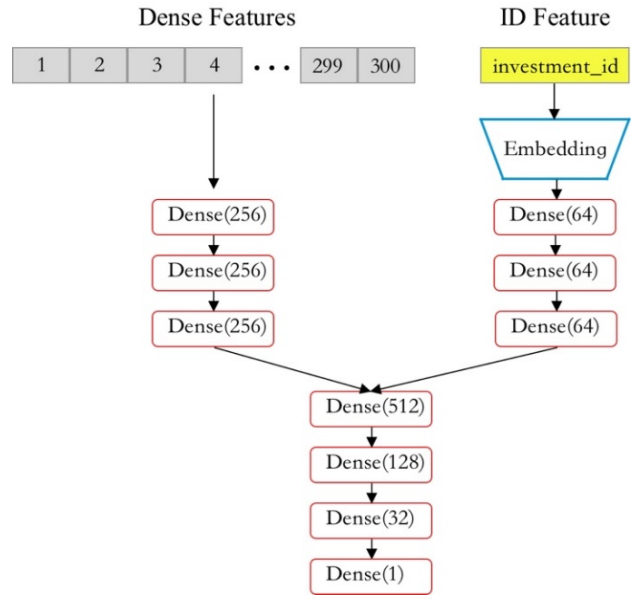

Figure 1. The Structures of dual branch network

Our approach has two input branches, the first branch re-encodes dense features., which utilizes a three-layer perceptron with 256 nodes in each layer. The simplest multilayer perceptron consists of an input layer, a hidden layer and an output layer. In order to obtain more robust features, we use three hidden layers to encode the input multiple times. The second branch also contains three-layer perceptron with 64 nodes in each layer. Before accessing the MLP, the investment id is encoded into a dense vector by embedding, so as to map the original high-dimensional data to the low-dimensional manifold. In this way, the original id features dimension can be compressed with a compression ratio more than 20. The low-dimensional manifold after embedding not only contains semantic information, but also simplifies the optimization process. Both branch networks are then connected to the same fully connected layer and combined together, the fully connected layer has a total of 512 nodes. Finally, the results are output through three fully connected layers. The number of nodes in the three fully connected layers is 128, 32 and 1 respectively.

Swish Layer leverages the same value for gating to simplify the gating mechanism, and no boundary on the right side avoids oversaturation. The smooth activation function makes the gradient return smoother, resulting in better accuracy and generalization.

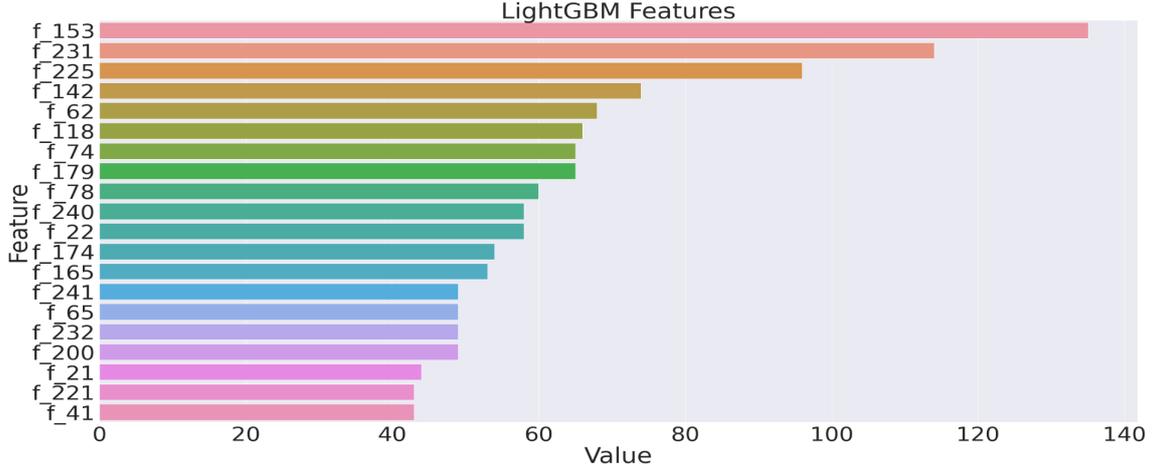

Figure 2. the importance of features

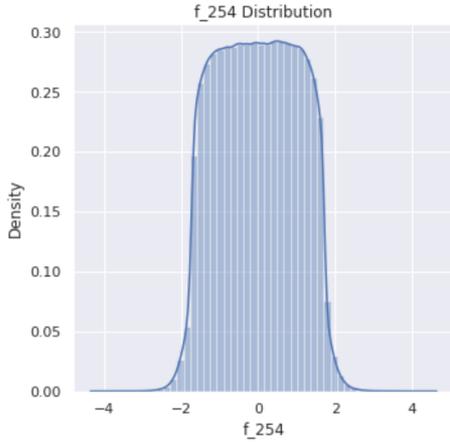

Figure 3. data distribution for f_254

## IV. EXPERIMENTS

The dataset utilized in this paper is the Ubiquant Market Prediction dataset published by Kaggle. In order to protect related privacy, most of the features have been desensitized. The partial desensitization data distribution is shown in Figure 3. In addition, we leveraged Lightgbm to rank the importance of features, as shown in Figure 2. The ranked features assist to analyze and mine the features required for the task more deeply. It can also be used to do feature selection. For example, the features f_253, f_231, f_225, f_142 and f_62 have relative high feature importance so that the memory size of input data can be reduced.

Table 1. Comparison of different algorithms on Pearson Coefficient

| Models | Pearson Coefficient |
|---|---|
| Xgboost | 0.137 |
| Lightgbm | 0.1382 |
| Catboost | 0.1365 |
| Ours | 0.1524 |

This paper utilizes Pearson Coefficient as the evaluation index. Pearson Coefficient ranges from -1 to +1, where 0 represents no correlation, negative values represent negative correlation, and positive values represent positive correlation. The Pearson correlation coefficient is optimized on the Euclidean distance, and the values of the vectors are centered, that is, the mean of the elements is subtracted for all dimensions in the two vectors. After centralization, the average value of all dimensions is basically 0, and then the cosine distance is calculated for the centralization result, as formula (1).

$$\rho_{X,Y} = \frac{\mathbb{E}[(X - \mu_X)(Y - \mu_Y)]}{\sigma_X \sigma_Y} \quad (1)$$

During training, the optimizer is Adam, the learning rate is set to 0.001, the batch size is 1024, and the epochs is 100. The Adam optimizer is an adaptive optimizer which can adjust the learning rate according to the current gradient and past gradient. The dual-branch adopts a step-by-step training method. First, a single branch is trained to make the weight of the single branch reach a better state, and then the two branches are combined for training. This training mode can maintain the feature difference of the two branches, making the fused features more robust. Besides, due to the rapid change of model weights and the limited mini-batch size at the beginning of training, the sample variance is large, so adopts the warm up strategy for training optimization. On the one hand, warm up helps to slow down the early overfitting of the mini-batch in the initial stage of the model and keep the distribution stable. On the other hand, it helps to maintain the stability of the deep layer of the model. The comparison results between our approach and the current popular algorithms such as Xgboost are shown in Table 1. Experimental results demonstrate that our approach achieves better performance than other approachs. Taking Pearson Coefficient as an indicator, our approach reached 0.1524, which was 1.54% higher than Xgboost, 1.42% higher than Lightgbm, and 1.59% higher than catboost. Compared with other approachs, our approach has done more detailed and effective work in feature encoding. Through the design of embedding and dual branch network structure, the network captures more fine-grained and effective features, so as to achieve better performance.

## V. CONCLUSIONS

In this paper, a novel dual branch network is proposed for the investment return prediction. The network structure takes the embedding extracted from investment id and the fine-grained features as input and predict the investment return as a regression task. The embedding method can reduce the input dimension with a compression ratio more than 20. Compared with popular methods such as xgboost and lightgbm, the approach proposed in this paper has different degrees of improvement in performance. In particular, it achieves a pearson coefficient of 0.1524, which is higher than the other two methods. In the future, we will investigate more feature engineering and improve the model structure to achieve a better result.

ACKNOWLEDGEMENT

Thanks to the Kaggle platform and Ubiquant Investment (Beijing) Co. for the dataset, which was instrumental in our research. Jianlong Zhu thanks Dan Xian for his idea contribution and feature engineering. Fengxiao and Yichen Nie are responsible for the expriments of this work. We also thank them for their coding part like model building, training and evaluation.